\documentclass[]{article}

\usepackage{arxiv}
\usepackage[utf8]{inputenc}
\usepackage{graphicx}
\usepackage{graphics}
\usepackage{authblk}
\usepackage[english]{babel}
\usepackage{csquotes}
\usepackage{mathtools}
\usepackage{comment}

\usepackage[backend=biber,style=ieee]{biblatex}
\usepackage[dvipsnames]{xcolor}
\usepackage{xparse}
\usepackage{authblk}
\usepackage{amsmath,amssymb,amsfonts,amsthm}
\usepackage[ruled,vlined,linesnumbered]{algorithm2e}
\usepackage{balance}
\usepackage{booktabs}
\usepackage{lipsum}
\usepackage{multirow}
\usepackage{subfig}
\usepackage{textcomp}
\usepackage{url}
\usepackage[dvipsnames]{xcolor}
\usepackage{setspace}
\usepackage{blindtext}
\usepackage{hyperref}

\usepackage{algpseudocode,float}

\usepackage{babel,blindtext}

\title{A Modified SEIR Model for the Spread of COVID-19 Considering Different Vaccine Types}
\author[]{Aram Ansary Ogholbake}
\author[]{Hana Khamfroush}

\affil{Department of Computer Science, University of Kentucky}

\begin{document}
\maketitle
\begin{abstract}
The COVID-19 pandemic has influenced the lives of people globally. In the past year many researchers have proposed different models and approaches to explore in what ways the spread of the disease could be mitigated. One of the models that have been used a great deal is the Susceptible-Exposed-Infectious-Recovered (SEIR) model. Some researchers have modified the traditional SEIR model, and proposed new versions of it. However, to the best of our knowledge, the state-of-the-art papers have not considered the effect of different vaccine types, meaning single shot and double shot vaccines, in their SEIR model.
In this paper, we propose a modified version of the SEIR model which takes into account the effect of different vaccine types. We compare how different policies for the administration of the vaccine can influence the rate at which people are exposed to the disease, get infected, recover, and pass away.
 Our results suggest that taking the double shot vaccine such as Pfizer-BioNTech and Moderna does a better job at mitigating the spread and fatality rate of the disease compared to the single shot vaccine, due to its higher efficacy.
\end{abstract}

\section{Introduction}
The COVID-19 pandemic has influenced all countries in the world. Each government has adopted different policies to mitigate the disease.
Lock downs, social distancing, masks, etc. were some of these policies that were put in place during the past year before the production of vaccines. 
After the emergence of the vaccines for the COVID-19 disease, each country used its own policies to vaccinate people based on different factors, such as availability of the vaccines in that country. Since two types of vaccines exist, single shot and double shot, it is important to know how effective each would be against the total spread of the disease. 

Since the start of the pandemic, many models have been proposed for the spread of the disease, many of which \cite{radulescu2020management, he2020seir, tang2020prediction,foy2021comparing} employ the Susceptible-Exposed-Infectious-Recovered (SEIR) model \cite{Aron1984}.
Some works have considered modified versions of SEIR to take into account people who have tested positive and people who are quarantined, in addition to those who are susceptible, exposed, infected and recovered. Other works have taken into account the effect of vaccine distribution on different age groups \cite{foy2021comparing}. 

To the best of our knowledge no previous work has considered the different types of vaccine (single shot and double shot) in their SEIR model. In this work, we take into account these different types of vaccine, and propose a new version of the SEIR model. The vaccines differ substantially from each other. The double-shot vaccine requires a waiting period, but provides high immunity, while the single dose vaccine has no waiting period, but its efficacy is lower.
Because these factors can influence both the spread of the disease and its fatality rate (the numbers of infected and deceased people),
in this paper we study how the choice of vaccination policy can affect the spread of the disease.

The structure of the rest of the paper is as follows: In section II, we review some of the previous work in modeling COVID-19. In Section III, we state the purpose of the paper and describe our proposed model. In section IV, we show our results, and finally in section V, we make our conclusion.

\section{Related Work}
There is a vast amount of work in modeling the COVID-19 disease. In this section, we summarize some of these works. We sub categorize these work into 3 groups: Graph Theory Model for COVID-19, No Vaccination SEIR Model For COVID-19, Vaccine Included SEIR For COVID-19.

\subsection{Graph Theory Model for COVID-19}
Graph theory has been used to model COVID-19. For example \cite{bhapkar2020virus} use a variable graph to model the disease.
Four graph types are defined in that work: Virus Graphs I to IV.  
COVID-19 was a type I Virus graph at the beginning, where people were divided into being either infected or not. The spread of corona virus could be stopped by putting all edges adjacent to active vertices of the infected group or all edges adjacent to the non-infected group in two disconnected sets. After the number of deaths due to the COVID-19 started increasing, its graph changed to type III. Now, due to the existence of the vaccine, COVID-19 has a type IV graph.

\subsection{ No Vaccination SEIR Model For COVID-19}
In \cite{unknown} the authors used a SEIR agent-based model for the COVID-19 spread. In their graph model, people are represented by nodes, and their contacts by edges. The nodes are categorized in the four groups of Susceptible, Exposed, Infected and Recovered. Also, they defined two additional groups for nodes which 1) have tested positive and 2) are quarantined. A proportion of the nodes at each simulation step is being tested and the infected and exposed nodes have a positive result. Nodes that have tested positive become quarantined. Since this paper aims to account for super spreaders which are the nodes with high degrees (many edges), they use a power law distribution for the degree distribution of nodes to be able to get more nodes with high degrees. However, in this work vaccination was not taken into account. In our paper, we have vaccinated states for the population.

In another paper \cite{1}, the authors have tested their model using a real-time data set. Using a SEIR model, they tracked the contacts and mobility of patients from 14 days before the start of initial symptoms until 7 days after it. Then they introduced a contact model for the contacts among people, and defined a probability for getting infected based on that contact model. The probability of being infected is based on two components: spatial and temporal. After running different strategies on their network, they showed that to optimally dismantle the giant connected components (GCC), it would be better to quarantine people who have high betweenness centrality. 

\subsection{Vaccine Included SEIR For COVID-19}
There are some papers that consider the vaccination in the SEIR model. This paper, \cite{foy2021comparing}, studies the effect of different vaccination strategies in India. They modified a conventional SEIR model to account for vaccination of different age groups. In the first strategy, the distribution of vaccine in the population is even. In the second strategy, the vaccine is first given to people of age 20-40. Strategies 3 and 4 give the vaccination first to the people of age 40-60, and more than 60, respectively. Their results show that all strategies have significantly lower infections compared to no vaccination. However, strategy 4 shows much lower cumulative deaths compared to the other strategies, and to no vaccination. Although their model considers the vaccine state, they don't consider the different vaccine types and how these vaccines could affect the spread of the disease. Moreover, in their model, only the susceptible people get vaccinated. However, in the real world scenario the recovered people from the disease can also get vaccinated due to the possibility of reinfection for them. In our work, both susceptible and recovered people can get vaccinated.

In another paper \cite{ghostine2021extended}, the authors proposed an extended SEIR model in which the susceptible could get vaccinated and then they can get exposed based on the efficacy of the vaccine. 

\section{Problem Statement}
The main goal of this project is to study the effects of different types of vaccination on the spread of the COVID-19 disease. Different types of vaccines exist against the disease. Some of them need only one dose while the others require two doses. Moreover, the latter require some time to pass between the two doses. Obviously, the spread of the disease can be affected by different vaccination policies (e.g. type of vaccine given, timing between doses, etc.) in different countries. Below we present our model, and explore how different policies can affect the spread of the disease. 

\begin{figure}[t]
\centering
\includegraphics[scale=0.27]{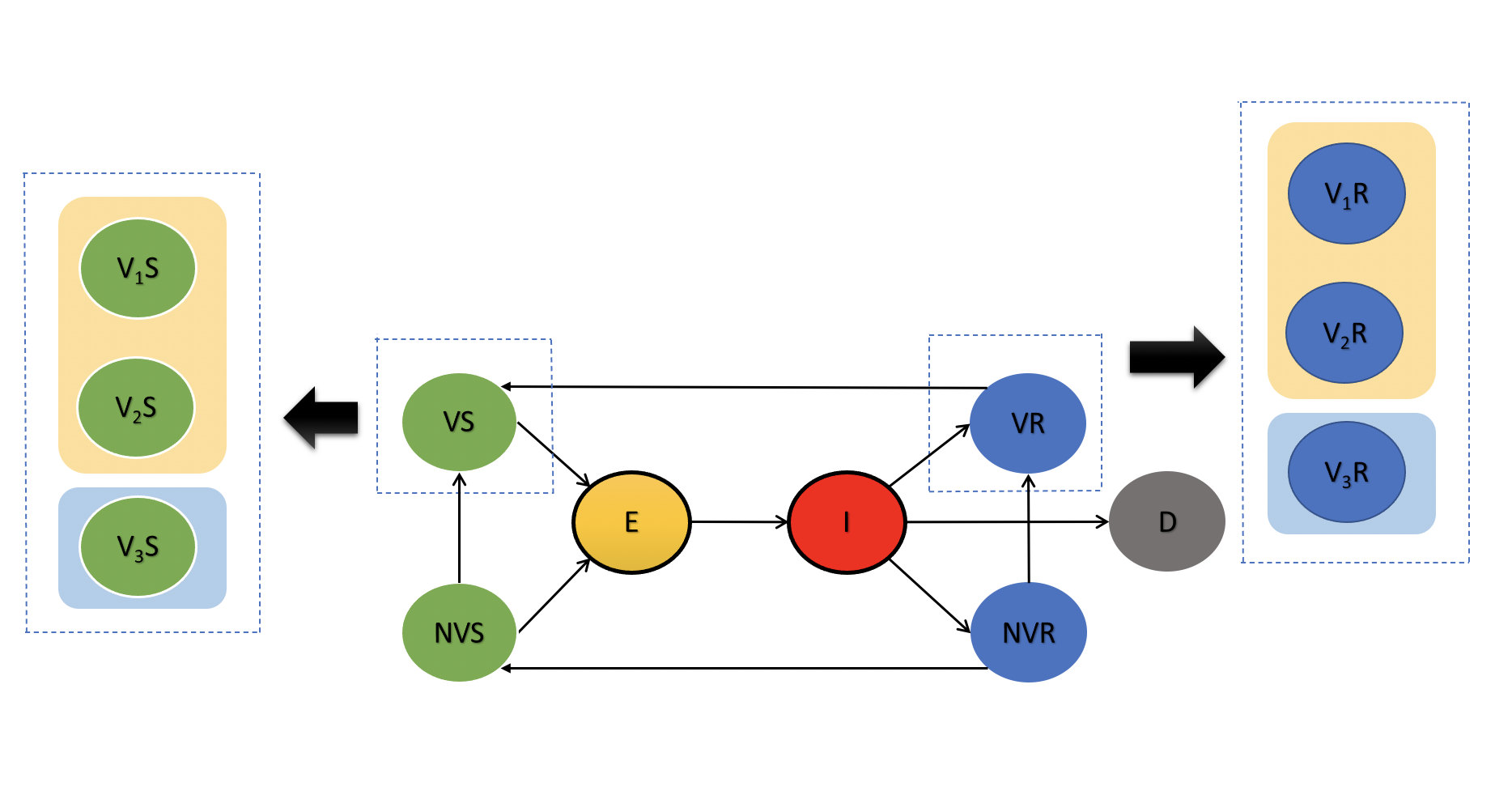}
\vspace{-2mm}
\caption{Proposed modified SEIRS Model}
\end{figure}

\subsection{Proposed Solution}
In this paper, we use a modified version of the SEIR model, called the SEIRS model, for modeling the spread of COVID-19. In the SEIR model \cite{unknown}, at any time individuals in the network can be in one of 4 different states: 1) Susceptible: these are individuals who are not infected; 2) Exposed: these are people who are in the incubation period; 3) Infected: these people are infected; and 4) Recovered: these individuals are either recovered from the disease or deceased.
 The SEIRS model \cite{trawicki2017deterministic} allows for reinfection.
Therefore, a fraction of recovered people can lose their immunity after some time, and get reinfected with COVID-19.

Since the purpose of this work is to study the effect of the vaccine on the population, we use the SEIRS model, but we also allow for susceptible and recovered individuals to become vaccinated.
As shown in Figure 1, the susceptible state is divided into two groups of Vaccinated Susceptible (VS) and Non-Vaccinated Susceptible (NVS). Moreover, the VS state itself contains three groups: $V_1S$, $V_2S$, and $V_3S$. $V_1S$ and $V_2S$, which are in a yellow box in Figure 1, are the susceptible people who took the first and second shots of the two dosage vaccine, respectively. $V_3S$, which is in the blue box shows the susceptible people who took the single shot vaccine.  The recovered state is the combination of Vaccinated Recovered (VR), Non-Vaccinated Recovered (NVR) and deceased people. By vaccinated recovered, we mean both groups of people who have taken the vaccine when they were susceptible but became infected after vaccination and the ones who take the vaccine when they recovered from the disease for the first time. Similarly to the VS group, the VR group is divided into recovered people who then take the single dose vaccine (blue box), or the double dose vaccine (yellow box). The double shot vaccine group is split into a group that has taken the first dosage of the vaccine, and another that has taken both dosages. Below is the detailed set of differential equations for our modified SEIRS model:
\begin{subequations}
\begin{align}
        &\frac{dNVS}{dt} = -\beta (NVS)I -\sigma NVS + \alpha_0 NVR \\  
        &\frac{dV_{1}S}{dt} = -\beta' (V_{1}S)I + \sigma_{V_{1}} NVS - \rho V_{1}S +\alpha_1 V_1R    \\
        &\frac{dV_{2}S}{dt} = -\beta'' (V_{2}S)I + \alpha_2 V_{2}R + \rho V_{1}S
\end{align}
\end{subequations}
\addtocounter{equation}{-1}
\vspace{-12pt}
\begin{subequations}
\setcounter{equation}{3}
\begin{align}
        &\frac{dV_{3}S}{dt} = -\beta''' (V_{3}S)I + (\sigma -\sigma_{V_{1}})NVS + \alpha_3 V_{3}R \\
                   &\frac{dE}{dt} = (\beta (NVS) +\beta'(V_{1}S) +\beta'' (V_{2}S) \nonumber \\
                   &\quad \quad \quad +\beta''' (V_{3}S))I - \epsilon E
        \\
        &\frac{dI}{dt} = \epsilon E - \gamma I\\
        &\frac{dV_{1}R}{dt} = \zeta_1 NVR - \alpha_1 V_1R +\gamma_{V_{1}} I - \iota V_1R\\
        &\frac{dV_2R}{dt} = \iota V_1R - \alpha_2 V_2R +\gamma_{V_2} I\\
        &\frac{dV_3R}{dt} = \zeta2 NVR - \alpha_3 V_3R +\gamma_{V_3} I\\
        &\frac{dNVR}{dt} = \gamma_{NV} I -\zeta NVR - \alpha_0 NVR\\
        &\frac{dD}{dt} = (\gamma-\gamma_V- \gamma_{NV})I
\end{align}
\end{subequations}
The symbols used in these equations are explained in Table~\ref{tab:symb}.

Equations (1a)--(1d) show the change in the rate of non-vaccinated and vaccinated susceptible people.
The rate of change of non-vaccinated susceptible individuals is due to: 1) contacts between non-vaccinated susceptible individuals with infected people, 2) non-vaccinated people getting vaccinated, and 3) non-vaccinated recovered people losing immunity and becoming non-vaccinated susceptible.
On the other hand, each group of the vaccinated susceptible state has its own rate of exposure to the disease.
In each group, the rates change with recovered people losing immunity and becoming susceptible again.
Some portion of non-vaccinated susceptible people takes the first dose of the double dosage vaccine, while another portion takes the single shot vaccine. Finally, some of those who took the first shot proceed to take the second.

Equation (1e) shows the change in the rate of exposed people. This rate is based on 1) susceptible people who become exposed, and 2) exposed people who become infected. Equation (1f) shows the change in the rate of infected state. This rate is affected by the 1) exposed people who become infected, 2) infected people who recover either to the vaccinated recovered state or the non-vaccinated recovered state, and 3) infected people who die from the disease.
\begin{figure}[b]
\centering
\includegraphics[scale=0.4]{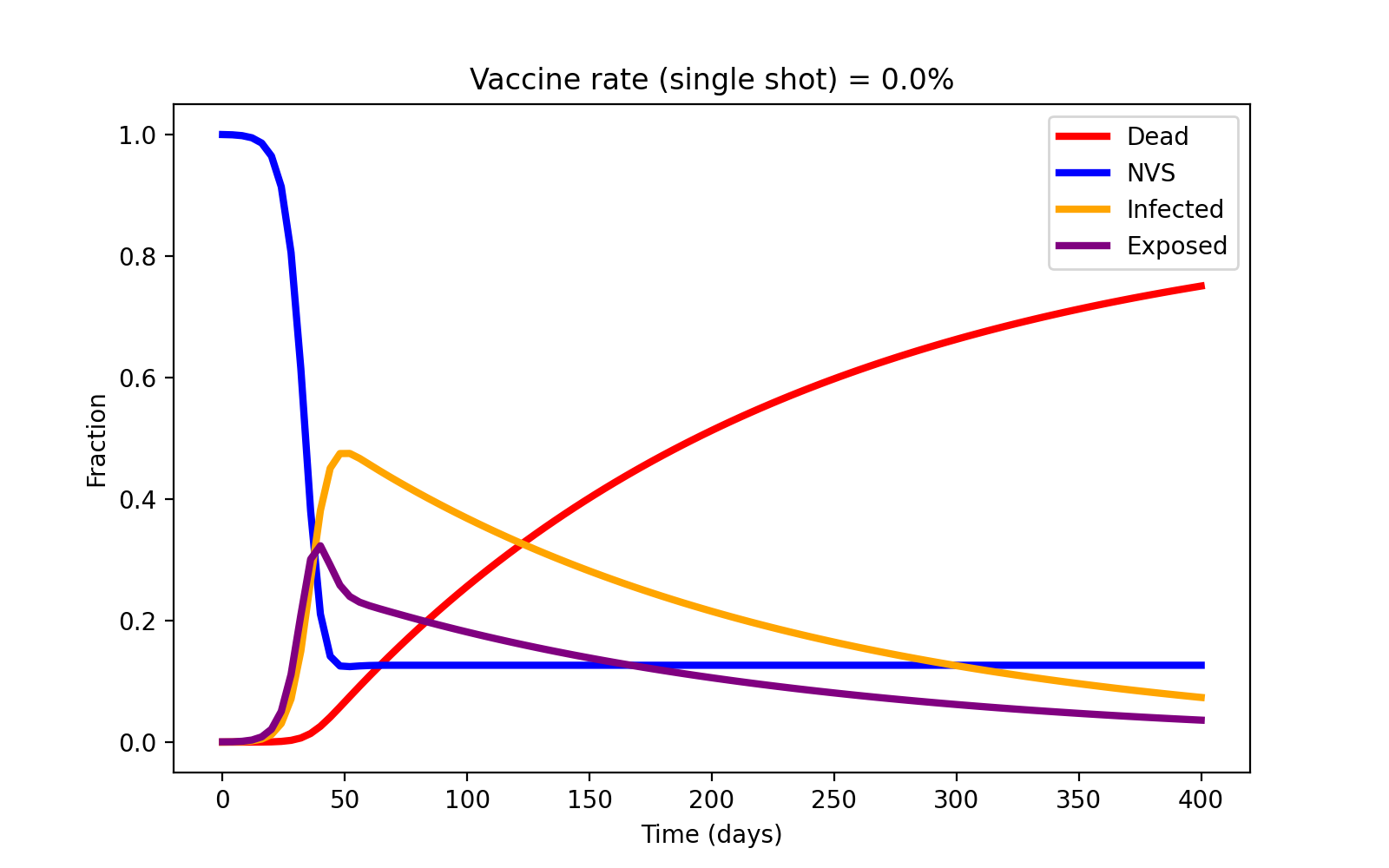}
\vspace{-2mm}
\caption{Fraction of population who are non-vaccinated susceptible, dead, infected and exposed over 400 days for no vaccine}
\end{figure}

\begin{figure}[b]
\centering
\includegraphics[scale=0.4]{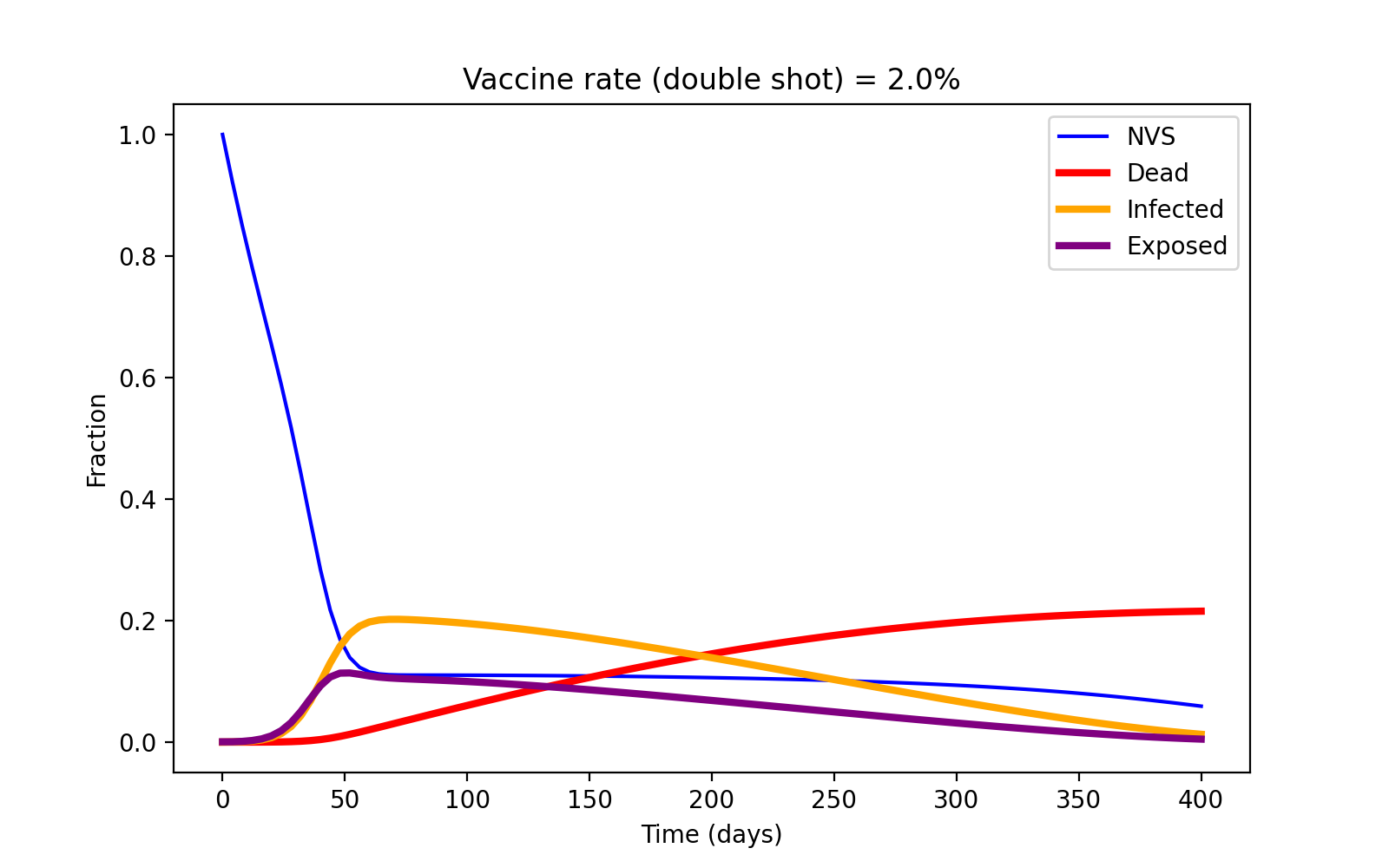}
\vspace{-2mm}
\caption{Fraction of population who are non-vaccinated susceptible, dead, infected and exposed over 400 days for double shot vaccine}
\end{figure}

\begin{figure}[t]
\centering
\includegraphics[scale=0.4]{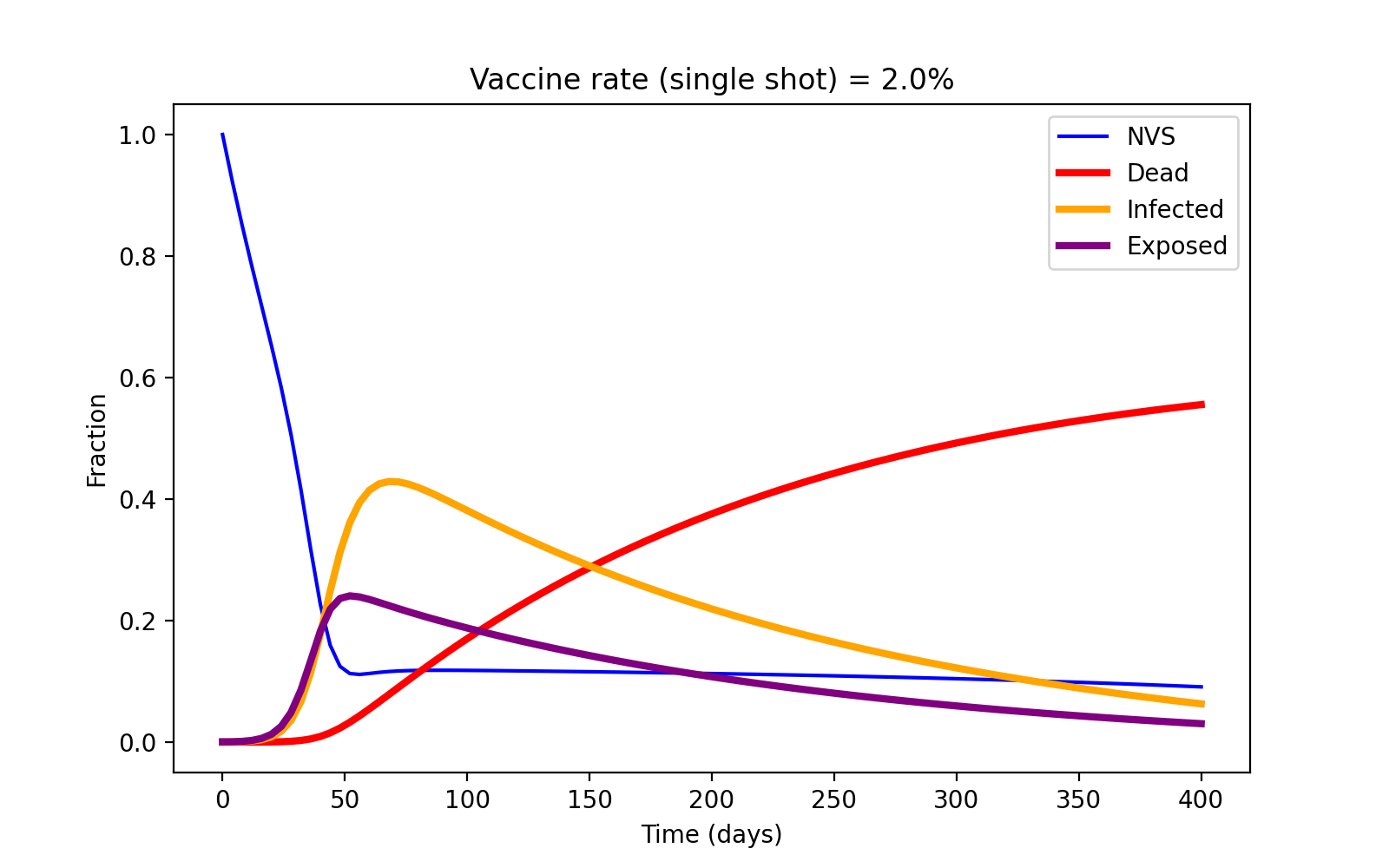}
\vspace{-2mm}
\caption{Fraction of population who are non-vaccinated susceptible, dead, infected and exposed over 400 days for single shot vaccine}
\end{figure}

\begin{figure}[t]
\centering
\includegraphics[scale=0.4]{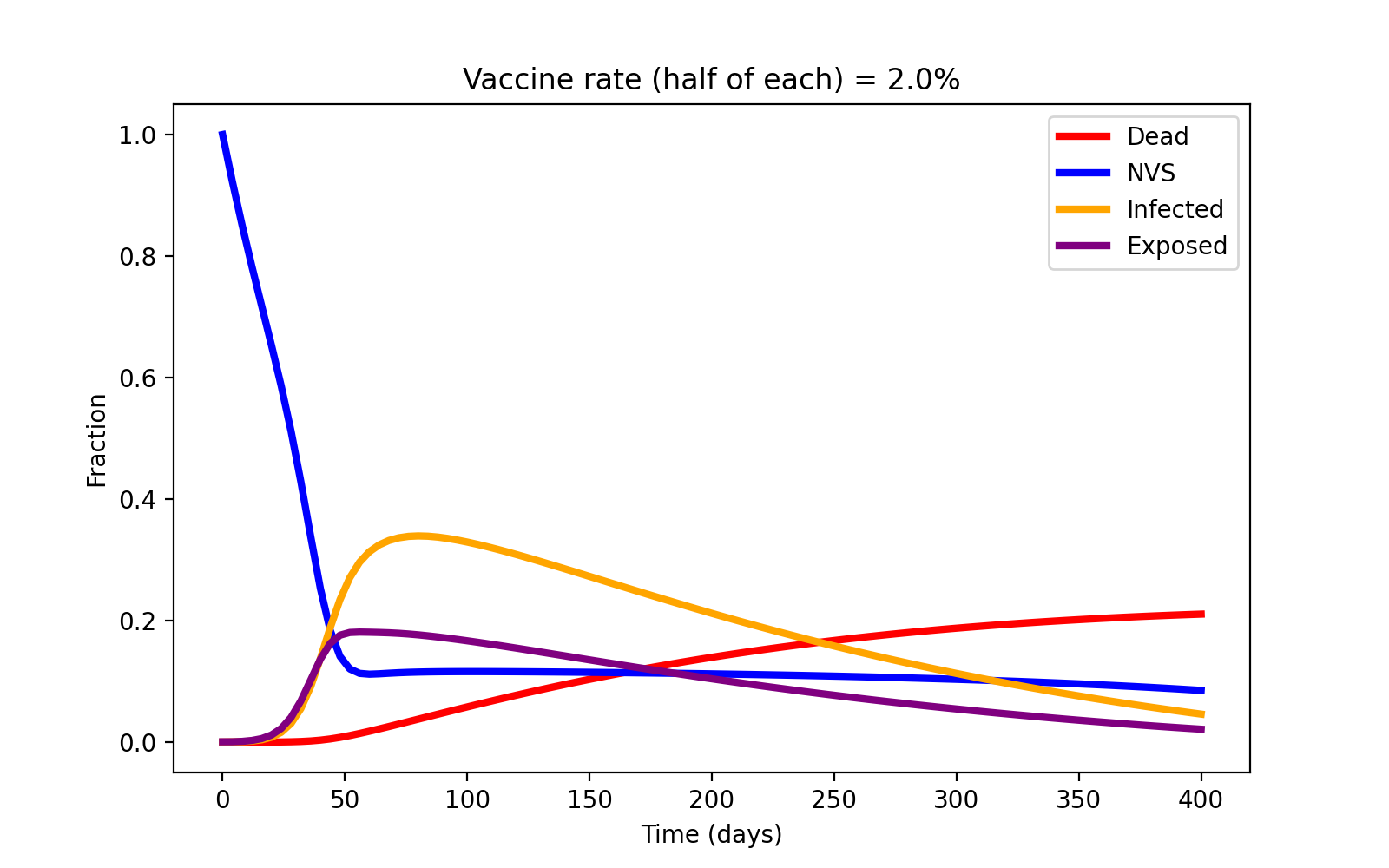}
\vspace{-2mm}
\caption{Fraction of population who are non-vaccinated susceptible, dead, infected and exposed over 400 days for half of each vaccine}
\end{figure}

 \begin{figure}[t]
\centering
\includegraphics[scale=0.4]{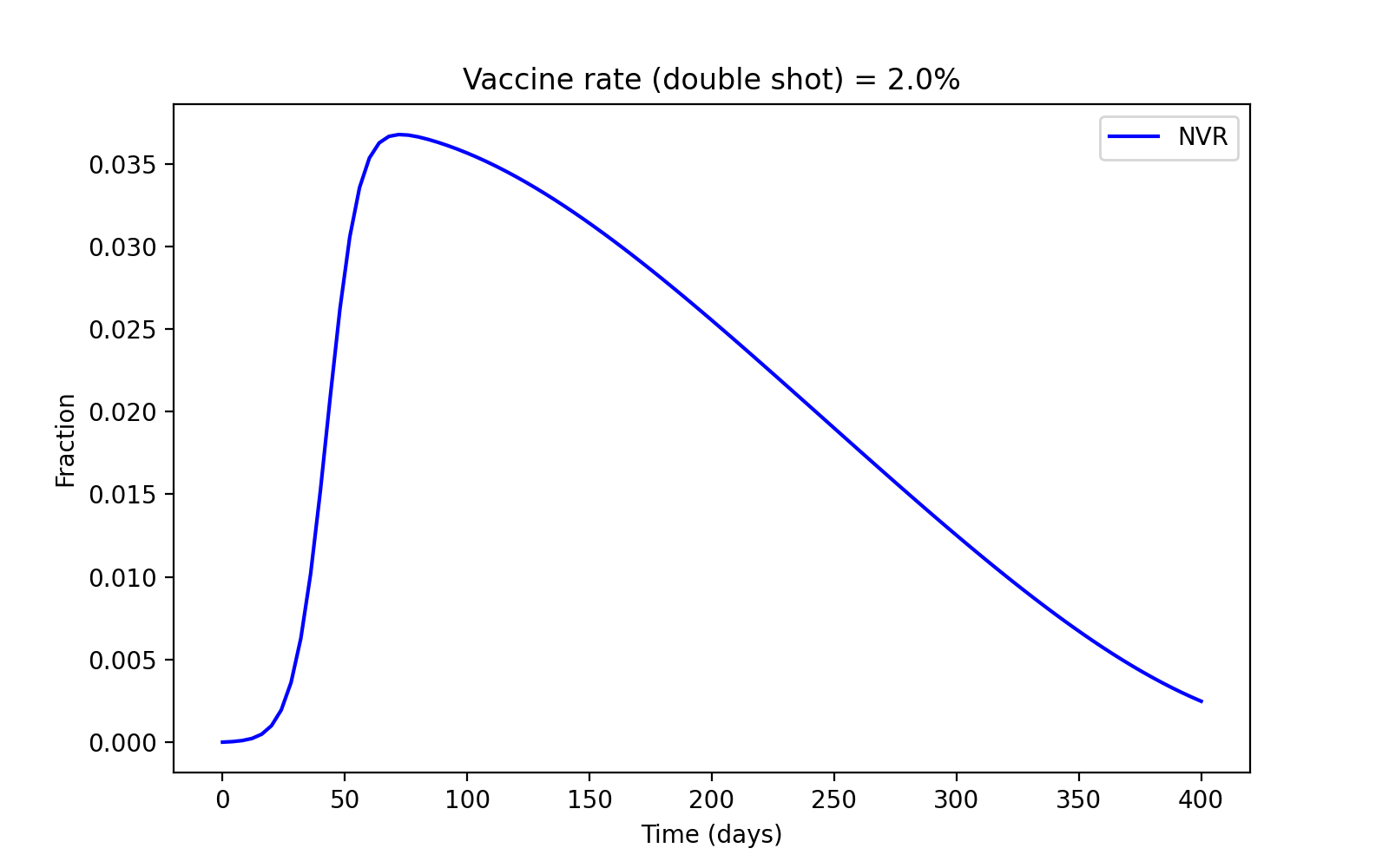}
\vspace{-2mm}
\caption{Fraction of population who are non-vaccinated recovered over 400 days for double shot vaccine}
\end{figure}

\begin{figure}[t]
\centering
\includegraphics[scale=0.4]{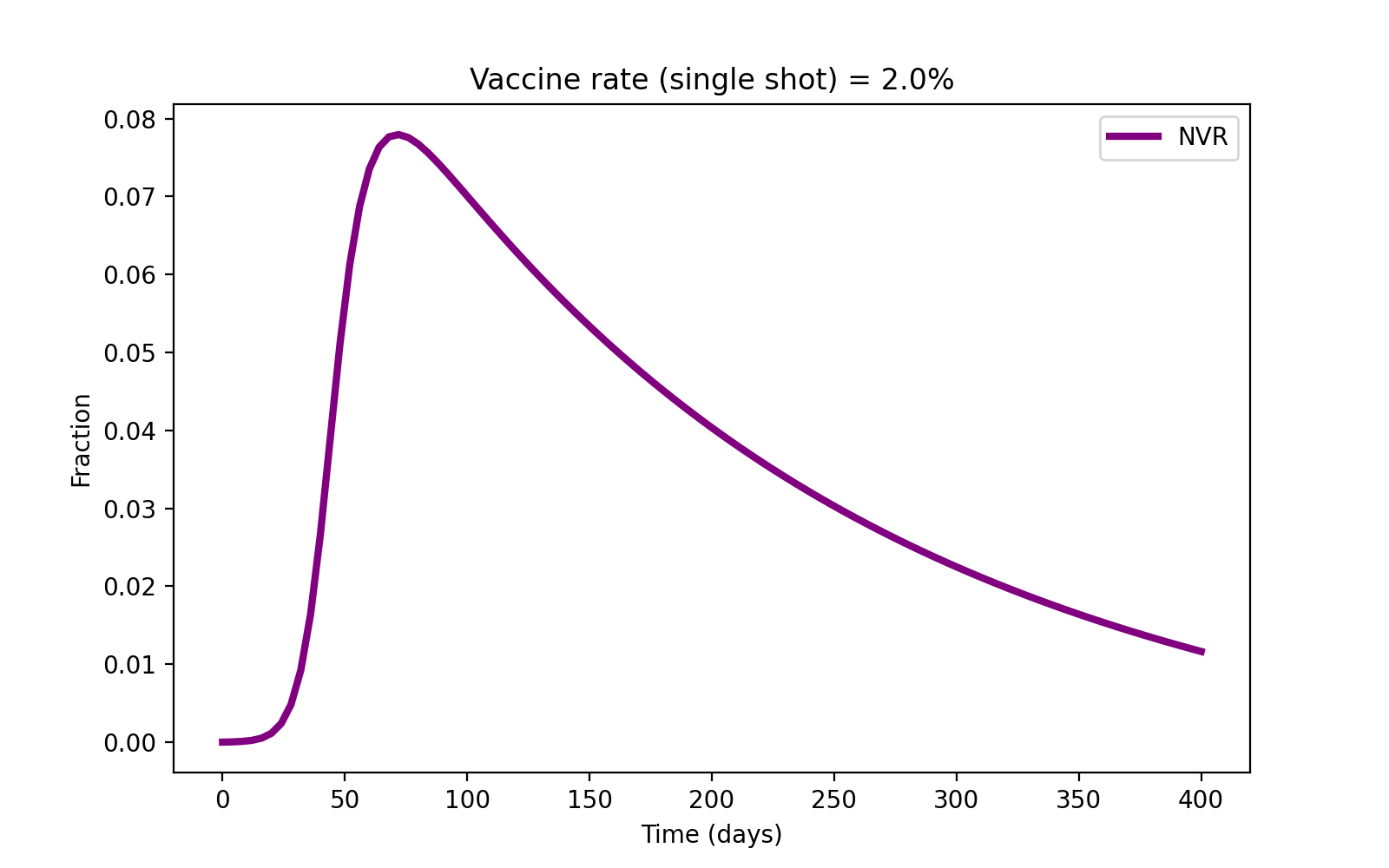}
\vspace{-2mm}
\caption{Fraction of population who are non-vaccinated recovered over 400 days for single shot vaccine}
\end{figure}

\begin{figure}[t]
\centering
\includegraphics[scale=0.51]{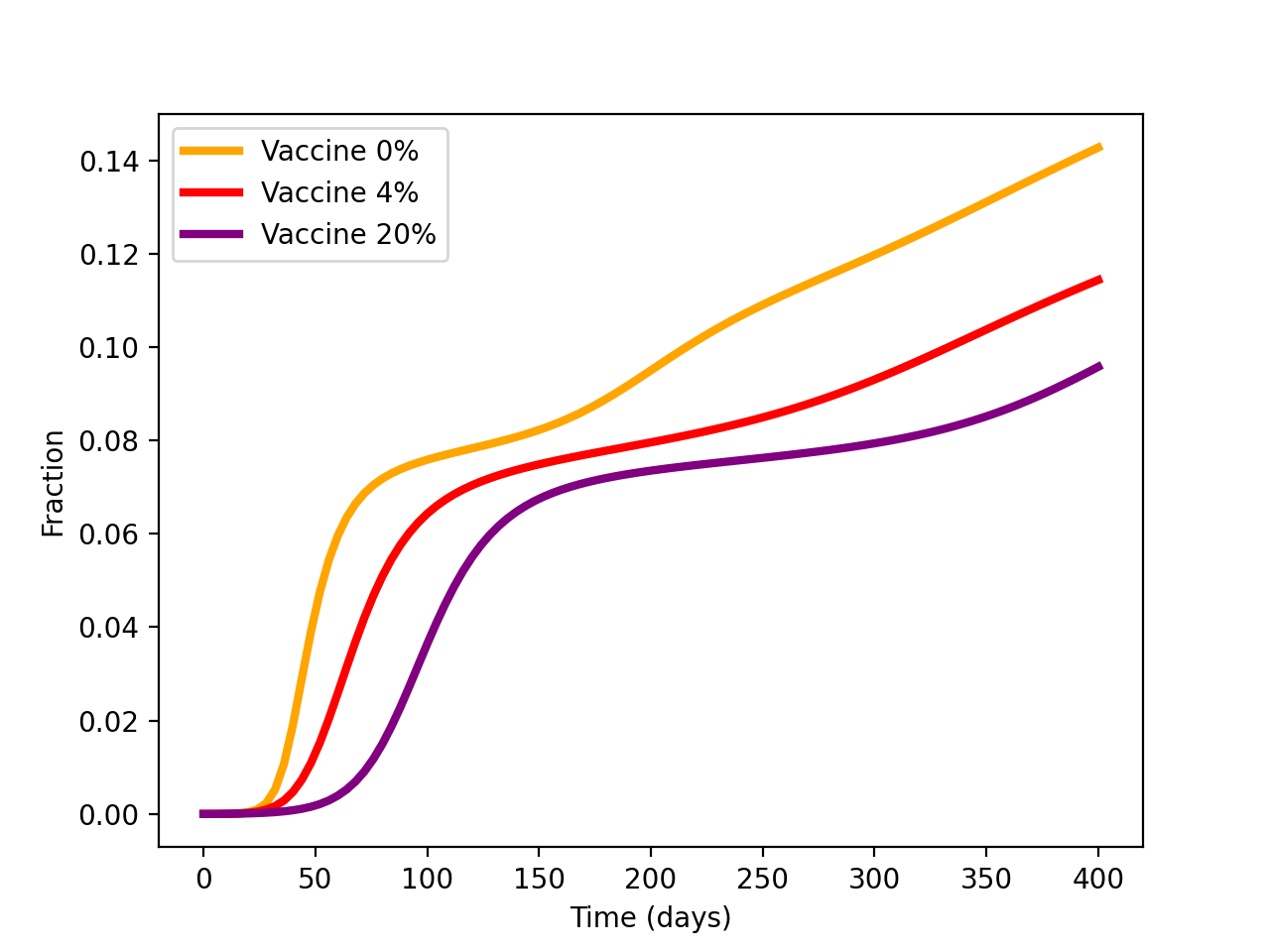}
\vspace{-2mm}
\caption{Fraction of population who are deceased for vaccination rates of 0\%-20\%}
\end{figure}

\begin{figure}[t]
\centering
\includegraphics[scale=0.51]{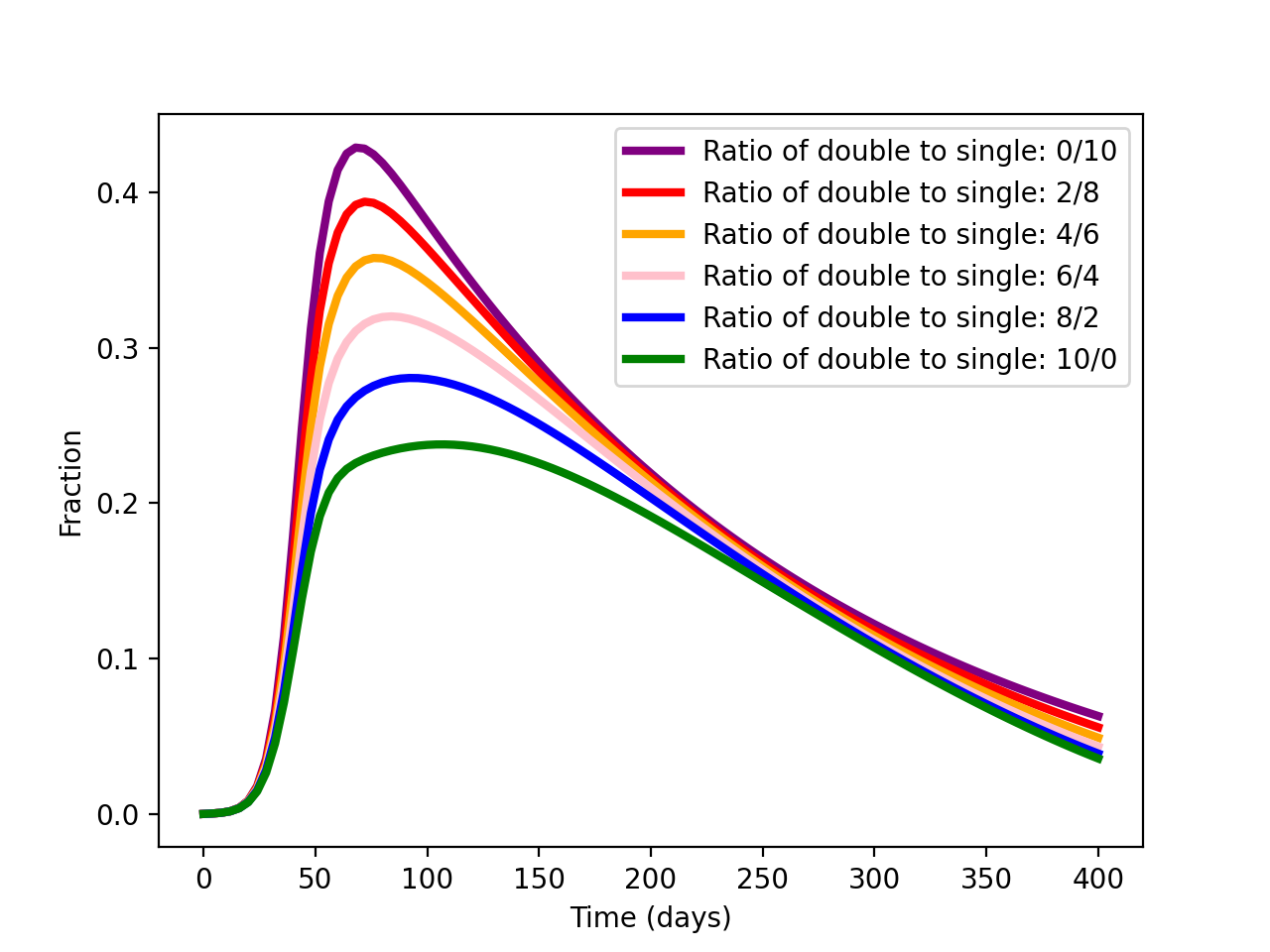}
\vspace{-2mm}
\caption{Fraction of population who are deceased for different vaccination proportions}
\end{figure}

Equations (1g)--(1j) show the rates at which infected people recover. For the non-vaccinated recovered group, the rate depends on 1) infected people who recover, 2) non-vaccinated recovered who take one of the vaccines, and 3) non-vaccinated recovered people who lose immunity and become susceptible. The rate of change in recovered people who take the first dose of the vaccine depends on non-vaccinated people who take the first shot of vaccine, infected people who recover, and those who take the second dose of the vaccine. Similar changes apply to the other groups of vaccinated recovered states. Exceptions are the rates for $V_2R$, who are not affected by non-vaccinated people, and $V_3S$ who are not affected by double-dose vaccine rates.

Equation (1k) is the change in the rate of deceased people.  This reflects on the portion of infected people with high risk of mortality.

\begin{table}[t]
\caption{State Transmission Parameters}
\label{tab:symb}
\centering
\begin{center}
{
 \begin{tabular}{||c c||} 
 \hline
 Definition & Parameter  \\ 
 \hline\hline
 rate of contacts & $\beta$ \\ 
 \hline
  rate of NVS become VS &  $\sigma$\\
 \hline
 rate of NVR become NVS & $\alpha_0$ \\  
  \hline
 rate of $V_1S$ become E & $\beta'$\\
 \hline
 rate of $V_2S$ become E & $\beta''$ \\
   \hline
 rate of $V_3S$ become E & $\beta'''$ \\
    \hline
 rate of NVS become $V_1S$ & $\sigma_{V_1}$ \\
     \hline
 rate of $V_1S$ become $V_2S$ & $\rho$ \\
      \hline
 rate of $V_1R$ become $V_1S$ &  $\alpha_1$\\
    \hline
 rate of $V_2R$ become $V_2S$ &  $\alpha_2$\\
    \hline
  rate of E become I & $\epsilon$\\
     \hline
 total rate of I become recovered & $\gamma$ \\
      \hline
 rate of I become NVR & $\gamma_{NV}$ \\
      \hline
rate of I become VR & $\gamma_{V}$ \\
      \hline
 rate of NVR become $V_1R$ & $\zeta_1$ \\
    \hline
 rate of NVR become $V_2R$ & $\zeta_2$ \\
 \hline
 rate of $V_1R$ become $V_2R$ & $\iota$ \\
 \hline
\end{tabular}
}
\end{center}
\end{table}

\begin{table}[t]
 \caption{ Set Up Values }
 \begin{center}
{
 \begin{tabular}{||c c ||} 
 \hline
 Parameter & Value  \\ [0.5 ex] 
 \hline\hline
 Population & 10000 \\ 
 \hline
 Incubation Period & 5.2 days\\  
  \hline
  Reproduction Number & 2.4 \\  
  \hline
Number of Initial infected & 5 \\ 
   \hline
Efficacy of Single Shot Vaccine & 0.65 \\
    \hline
Efficacy of Double Shot Vaccine & 0.95 \\
    \hline
\end{tabular}
}

\end{center}
\end{table}

\section{Experimental Results}
  To evaluate our model, we solved the differential equations of Section III \color{blue} \href{https://apmonitor.com/pdc/index.php/Main/SimulateCOVID19}{using this resource}\color{black}, (which is a COVID-19 SEIR simulator),
  for a population of 10,000 people in which only 5 are infected initially and the rest are in the state of non-vaccinated susceptible.
  Some  of  the  values used in our simulations are  shown  in  Table  II.
  We compare four different scenarios. First, we consider a scenario of no vaccinations are available. In the second scenario, we consider the vaccine rate of 2\% for double dose vaccines and the efficacy of 65\% after the first dose and the efficacy of 95\% after the second dose. In the third test, we consider the same rate of 2\% but for single-dose vaccines in which their efficacy is assumed to be 65\%. And at last, we consider a scenario in which the vaccine is half single shot and half double shot and the total amount is 2\%. Still, the single shot vaccine has 65\% efficacy while the full double shot vaccines are 95\% effective.
  
  As shown in Figures 2, 3, 4 and 5 when no vaccine is available, the mortality rate goes up to 80\%. But, with the double shot vaccines, the mortality fraction starts to increase after 50 days, and after 300 days it starts to flatten out at 20\%. However, using only single shot vaccine with the same rate of 2\%, the mortality is significantly higher than double shot vaccines, and reaches 60\% by 400 days, without flattening out. Moreover, the peak of the fraction of people in infected and the exposed states are almost twice for the case of single-shot vaccines compared to the double-shot vaccines. Also, based on Figure 5, we see the death rate is not very different compared to when only double-shot vaccines are given to the population. However, the fraction of infected and exposed population is still bigger than the scenario of giving double-shot vaccines to the population. The peak for the fraction of population who become infected is for the case of giving half single shot and half double shot is almost 40\% which is roughly twice as much as the scenario of only giving double-shot vaccines.
  
  Also, we compared the recovered fractions for the cases of using double shot vaccines and single shot vaccines. As shown in Figures 6 and 7, the fraction of non-vaccinated people who recover from the disease in the case of double shot vaccines is lower than the fraction of none-vaccinated people who recover in the case of the single shot. The reason is that the infection rate is lower in the case of double shot vaccines and not many people become ill and consequently they don't have any disease to recover from it.
 
 Moreover, we compared the effect of different rates of vaccines in a population on a fraction of deceased people for three different cases of no vaccination, 4\% vaccination and 20\% vaccination. As it's shown in Figure 8, our results show that as the vaccination rate increases the amount of deceased people decreases significantly.
 
 At last, we looked at different ratios of the double shot to single shot vaccines to be given to a population. Our results show that the different ratios of vaccine types almost have the same time of pick for the number of deceased people. However, the least amount of deceased people comes from the scenario where the whole population takes double shot vaccines.
 
\section{Conclusion}
With the emergence of the COVID-19 disease many researchers modeled the spread of the disease to be able to find an efficient way to mitigate its fatality and infection rate. The SEIR model is one of the most used models for that matter. However, to the best of our knowledge, no other work considered the different vaccine types in their SEIR models.

In this work, we proposed our modified version of the SEIR model which takes the different vaccine types, namely single shot and double shot vaccines, into account. We tested our model and showed that with the same rate of vaccination, the double shot vaccine would be a better choice for vaccine administration. Based on our results, the mortality rate of the population would decrease by more than half and the infected and exposed fraction of people would decrease significantly using double shot vaccines compared to single shot ones.
For future work, we plan to use more realistic numbers for the transmission rates than described in this work.

\printbibliography
\end{document}